\documentclass{article}
\renewcommand{\epsilon}{\varepsilon}
\begin{document}
\title{Parametric phenomena of the particle dynamics in a
periodic gravitational wave field}
\author{Alexander B. Balakin\footnote{Electronic address:
Alexander.Balakin@ksu.ru}\ ,
Veronika R. Kurbanova\footnote{Electronic address:
Veronika.Kurbanova@ksu.ru}\\
Department of General Relativity and Gravitation\\
Kazan State University, 420008 Kazan, Russia\\
and\\
Winfried Zimdahl\footnote{Electronic address: zimdahl@thp.uni-koeln.de}\\
Fachbereich Physik, Universit\"at Konstanz\\ D-78457 Konstanz, Germany\\
and\\
Institut f\"ur Theoretische Physik, Universit\"at zu K\"oln\\
D-50937 K\"oln, Germany}
\date{\today}
\maketitle
\begin{abstract}
We establish exactly solvable models for the motion of
neutral particles,
electrically charged point and spin particles (U(1) symmetry),
isospin particles (SU(2) symmetry), and particles with color charges (SU(3)
symmetry) in a gravitational wave background.
Special attention is devoted to parametric effects induced
by the gravitational field.
In particular, we discuss parametric instabilities of the particle motion
and parametric oscillations of the vectors of spin, isospin, and color charge.
\end{abstract}
\vspace{1.5cm}
\section{Introduction}
\label{introduction}

Periodic external fields are known to induce parametric phenomena in physical systems.
This includes such effects as  parametric
oscillations (the oscillation frequency becomes a periodic function of time)
and parametric instabilities (exponential growth of certain dynamical
quantities) \cite{Cap,Stoker,Arscott}.
Classical examples are the parametric resonance in vibrations of
mechanical and electrical systems \cite{Cap,Stoker,Arscott} and plasma
instabilities in  external electromagnetic fields \cite{Jackson,Silin}.
From a mathematical point of view, such phenomena are described by differential
equations with periodic coefficients which are subject to Floquet's theory
\cite{Stoker,Arscott,Floquet,McLa,YaSta,Meix}.
As we shall demonstrate here, equations of
this type naturally appear if one considers the motion of different kinds of
particles in a gravitational wave (GW) background.
The gravitational wave  may
play the role of an external periodic pumping field.
In fact, most of the attempts for a direct detection of gravitational waves
rely on this
concept. The mathematical similarity to dynamical equations which are known
to describe parametric oscillations and resonances then naturally suggests the
possibility of gravitationally induced parametric effects.

The idea of parametric
phenomena in GW fields is not new. For a linearized GW field it was
discussed in   \cite{GribFr}.
The first exactly solvable model for
the evolution of a kinetic system in a {\it nonlinear} GW field, demonstrating
explicitly the possibility of parametric excitation of a
relativistic plasma by a
periodic GW, has been established in \cite{Bal}.
At the same time, in the 90th, the problem of parametric resonance during
the reheating phase of inflationary models has become an intensely
elaborated topic in a
cosmological context (see, e.g., \cite{Tra,Kof1,Kof2,Fin,Taru}).
More recently, investigations, concerning parametric phenomena in a
GW field, have attracted attention again (see, e.g., \cite{KVP,vanH,BM}).

The purpose of the present paper is to clarify characteristic
features of gravitationally induced parametric effects for simple
dynamical configurations. As a first example we consider the
motion of an electrically charged point particle which is
simultaneously exposed to a constant magnetic field and a
gravitational wave with front plane orthogonal to the magnetic
field. In the second example we include an additional spin degree
of freedom which is described by the Bargmann-Michel-Telegdi (BMT)
equations \cite{BMT}. In the third case the electrically charged
spin-particle is replaced by a particle with isospin and the
magnetic field is replaced by a corresponding Yang-Mills field.
The isospin dynamics is governed by Wong's equation \cite{Wong}
for the three-dimensional isospin vector. Finally, we consider the
motion of particles with color charge, described by Wong's
equation for the eight color degrees of freedom. Using suitably
specified Yang-Mills fields, we establish a general scheme which
allows a unified treatment of the particle dynamics for all four
cases. We show that on a periodic GW background this dynamics is
characterized by Hill and Mathieu equations. Well-known stability
properties of the latter allow us to classify the particle motion
accordingly. This implies parametric oscillations and/or
parametric instabilities as generic phenomena. The precession
dynamics of the vectors of spin, isospin and color charge is
coupled to the particle motion and parametrically driven as well.

The paper is organized as follows.
In Section \ref{particle dynamics} we establish the basic dynamic
equations for the Abelian
and non-Abelian subcases to be discussed in the following.
In Section  \ref{Yang-Mills}  the Yang-Mills fields for the latter cases are specified.
The (exact) gravitational background is characterized in Section \ref{GW}.
Section \ref{solution} is devoted to a compact, general solution for the particle dynamics.
A ``sandwich'' GW is considered as a special case.
The spin precession for an electrically charged particle is the subject of Section  \ref{spin}.
Sections \ref{isospin} and \ref{color} discuss the dynamics in the spaces of isospin and color charge, respectively.
In Section \ref{conclusions} we summarize our main results.
We use units in which $\hbar = c = 1$.

\section{Particle dynamics: Basic equations}
\label{particle dynamics}

Let us consider the
evolution of relativistic point particles with either an electric charge
and a spin-vector, or an isospin, or a color charge.
The concepts of classical particles with isospin (for the SU(2) symmetry) or
color charge (for the SU(3) symmetry) are generalizations of the electrically charged
particles to the non-Abelian case (see, e.g., \cite{Litim}).
The dynamical equations for the
particle momentum $p^{i}$,
for the spin-vector  $S^{i}$
and for  the charge
$Q^{\left( A \right)}$,
where $\left( A \right)$ is a group index, are
\begin{equation}
\frac{\mbox{D} p^i}{\mbox{D}\tau} = {\cal F}^i \ , \quad
\frac{\mbox{D}S^i}{\mbox{D}\tau} = {\cal G}^i \  \quad {\rm and}\quad
\frac{\mbox{D}Q^{(A)}}{\mbox{D}\tau} = {\cal G}^{(A)}\ ,
\label{1}
\end{equation}
respectively.
Here, $\mbox{D}$ denotes the covariant differential, $\tau$ is a parameter along the
particle worldline and  ${\cal F}^i$ is the
force four vector which is orthogonal to the particle momentum
$p^i = m \mbox{d}x^i/\mbox{d}\tau$, i.e.,
$p_i {\cal F}^i=0$.
The quantity ${\cal G}^i$ describes the spin
rotation within the BMT theory.
The quantity ${\cal G}^{(A)}$ is a vector in the group space which determines
the non-Abelian charge evolution and plays a similar role for the charge as ${\cal F}^i$
plays for the momentum.
The limiting case of neutral particles is
characterized by
$ {\cal F}^i = {\cal G}^i = {\cal G}^{(A)} = 0 $.

\subsection{Electrically charged point particles}

In this case the relevant force is the Lorentz force
\begin{equation}
{\cal F}^i = \frac{e}{m}F^i_{ \cdot k} \ p^{k}
\label{2}
\end{equation}
with
\begin{equation}
F_{ik}= \nabla_{i}A_{k} - \nabla_{k}A_{i}, \quad
\nabla_{k}F^{ik} = 0 \ ,
\label{3}
\end{equation}
where $F _{ik}$ is the Maxwell tensor.
The particle under consideration is regarded here as a test particle.
Moreover, the charge is constant which renders the third equation in
(\ref{1}) irrelevant.

\subsection{Electrically charged spin particles}

According to \cite{BMT} the
evolution of classical relativistic spin particles is governed by the
equations
\begin{equation}
\frac{\mbox{D}{p^{i}}}{\mbox{D}\tau} = \frac{e}{m} F^i_{ \cdot k} \ p^{k}
\label{4}
\end{equation}
and
\begin{equation}
\frac{\mbox{D}{S^{i}}}{\mbox{D}\tau} = \frac{e}{2 m}
\left[ \ g  F^i_{ \cdot k} \
S^{k} + \frac{(g - 2)}{ m^2}  p^i F_{kl} S^k p^l \right]\ .
\label{5}
\end{equation}
Here, $S^i$ is the spin four-vector and
$g$ is the gyromagnetic ratio. Equation (\ref{5}) describes the precession
of the magnetic moment.
It generalizes earlier non-relativistic equations by Thomas \cite{Thomas} and
Bloch \cite{Bloch}, which rely on the
circumstance that the ``expectation value of the vector operator
representing the ``spin" will necessarily follow the same time
dependence as one would obtain from a classical equation of
motion" (cf. \cite{BMT}).
While the particle momentum according to Eq. (\ref{4})
is independent of the spin vector, the dynamics of the latter is coupled (at
least via the covariant derivative) to the particle motion.

\subsection{Isospin particles}

Here we have a triplet $I^{(A)}$ of scalar
fields representing a vector in the three-dimensional isospin space, i.e.,
$(A) = (1), (2), (3)$.
This space has an euclidean metric $G_{(A)(B)}$.
The relevant force
\begin{equation}
{\cal F}^i=\frac{g}{m}F^{(A)i}_{ \ \ \ \cdot \ k} \ p^{k}I^{(B)} G_{(A)(B)}
\ ,
\label{6}
\end{equation}
where $g$ is the interaction constant,
has been obtained by Kerner \cite{Kerner} and Wong \cite{Wong}.
The isospin dynamics is determined by Wong's equation \cite{Wong}
\begin{equation}
\frac{\mbox{D}}{\mbox{D}\tau} I^{(A)}
= - \frac{g}{m}\epsilon^{(A)}_{\ \cdot \
(B)(C)}  A^{(B)}_{i} p^{i} \ I^{(C)}\ ,
\label{7}
\end{equation}
where we have used that the structure constants
for the $SU(2)$ group coincide with the three-dimensional Levi-Civita symbol
$\epsilon_{ \ \cdot \ (B)(C)}^{(A)}$.
The quantities $A_i^{\left(A \right)}$ are the vector potentials in terms of
which the Yang-Mills field strength tensor
 $F _{jk}^{\left(B \right)}$ is
given by
\begin{equation}
F^{(B)}_{jk}=\nabla_{j}A^{(B)}_{k}-\nabla_{k}A^{(B)}_{j}+
g \epsilon ^{(B)}_{\ \cdot \ (K)(L)}A^{(K)}_{j}A^{(L)}_{k} \ .
\label{8}
\end{equation}
The Yang-Mills field equations are
\begin{equation}
g^{ij}\left[\nabla_{i}F_{jk}^{(A)} + g\epsilon ^{(A)}_{\ \cdot \ (B)(C)}A_{i}^{(B)}F^{(C)}_{jk}\right]=0 \ .
\label{9}
\end{equation}
Again we consider the particle motion in a given external field.
Wong's equation represents a non-Abelian generalization of the
equation of motion for electrically charged point particles.
It can be obtained as the classical limit from quantum field
theory for the case of sufficiently localized quantum states of
the matter fields with characteristic length scales much smaller
than those associated to the Yang-Mills fields \cite{BW,Litim}.

\subsection{Particles with color charge}

For test particles with color charge the force ${\cal F}^i$ is given by
\begin{equation}
{\cal F}^i=\frac{g}{m}F^{(A)i}_{ \ \ \ \cdot \
k} \ p^{k}Q^{(B)} G_{(A)(B)}\ ,
\label{10}
\end{equation}
with the field strength tensor
\begin{equation}
F^{(B)}_{jk}=\nabla_{j}A^{(B)}_{k}-\nabla_{k}A^{(B)}_{j}+
g f^{(B)}_{\ \cdot \ (K)(L)}A^{(K)}_{j}A^{(L)}_{k}\ ,
\label{11}
\end{equation}
where the $f^{(B)}_{\ \cdot \ (K)(L)}$ are the structure constants of the SU(3)group.
The field equations are
\begin{equation}
g^{ij}\left[\nabla_{i}F_{jk}^{(A)} + g f ^{(A)}_{\ \cdot \
(B)(C)}A_{i}^{(B)}F^{(C)}_{jk}\right]=0 \ .
\label{12}
\end{equation}
The quantity $Q ^{\left(A \right)}$ is the color charge with
$\left(A \right) = \left(1 \right).... \left(8 \right)$.
Wong's equation in this case reads
\begin{equation}
\frac{\mbox{D}}{\mbox{D}\tau} Q^{(A)} = - \frac{g}{m}f^{(A)}_{\ \cdot \ (B)(C)}
A^{(B)}_{i} p^{i} \ Q^{(C)}\ .
\label{13}
\end{equation}
The structure constants  $f^{(A)}_{\ \cdot \ (B)(C)}$ are characterized by the commutator relations
\begin{equation}
\left[\lambda _{\left(A \right)}, \lambda _{\left(B \right)} \right]
= 2i f _{\left(A \right)\left(B \right)\left(C \right)}
\lambda ^{\left(C \right)} \,,
\label{14}
\end{equation}
where $\lambda_{\left(A \right)}$  are the traceless, hermitian Gell-Mann
matrices (see, e.g., \cite{Huang,Mosel}).
In detail we have
\begin{eqnarray}
f_{(1)(2)(3)}=1\ ,&&
f_{(4)(5)(8)}=f_{(6)(7)(8)}=\frac{\sqrt{3}}{2}\ ,\nonumber\\
f_{(1)(4)(7)}{=}f_{(2)(4)(6)}{=}f_{(2)(5)(7)}&=& f_{(3)(4)(5)}{=}
{-}f_{(3)(6)(7)}{=} {-}f_{(1)(5)(6)} {=} \frac{1}{2}\ . \nonumber\\
\label{15}
\end{eqnarray}
Below we shall also use the completely symmetric coefficients
$d _{\left(A \right)\left(B \right)\left(C \right)}$ of the basic
representation which are given by the anti-commutation relations
\begin{equation}
\left\{\lambda _{\left(A \right)}, \lambda _{\left(B \right)} \right\}
= \frac{4}{3}\delta _{\left(A \right)\left(B \right)}
+ 2 d _{\left(A \right)\left(B \right)\left(C \right)}
\lambda ^{\left(C \right)} \,,
\label{16}
\end{equation}
where \cite{ElseHeinz,Litim}
$$
d_{(1)(4)(6)} = d_{(1)(5)(7)} = d_{(2)(5)(6)} = d_{(3)(4)(4)} = d_{(3)(5)(5)}
=
$$
$$
= - d_{(2)(4)(7)} = - d_{(3)(6)(6)}  = - d_{(3)(7)(7)} = \frac{1}{2} \ ,
$$
$$
d_{(1)(1)(8)} = d_{(2)(2)(8)} = d_{(3)(3)(8)} = -
d_{(8)(8)(8)} =
 - 2 d_{(4)(4)(8)} =
$$
\begin{equation}
- 2 d_{(5)(5)(8)} = - 2 d_{(6)(6)(8)} = -
2 d_{(7)(7)(8)}
 = \frac{1}{\sqrt3}\ .
\label{17}
\end{equation}

\section{Yang-Mills fields with ``parallel'' potentials}
\label{Yang-Mills}

It is known that for each solution of the general relativistic
source free Maxwell equations one can construct a set of solutions
of the general relativistic Yang-Mills equations \cite{Yasskin}.
Following Gal'tsov \cite{Gal'tsov} we will refer to the
corresponding Yang-Mills potentials as ``parallel'' potentials.
The latter are characterized by
\begin{equation}
A^{(B)}_{i}=q^{(B)}A_{i}\ ,\quad
F^{(A)}_{ik} = q^{(A)} F_{ik}\ , \quad
q^{(B)}q_{(B)}=1\ , \quad q^{(B)} = {\rm const}\ .
\label{18}
\end{equation}
Due to the antisymmetry of the structure coefficients the
relations (\ref{8}) and (\ref{11})  as well as the equations (\ref{9}) and
(\ref{12}) reduce to the linear
Maxwell-type forms
(\ref{3}). Nevertheless, compared with Maxwell's  theory there exists an
additional
degree of freedom, namely the direction of the vector $q^{(A)}$ in the
group space \cite{Yasskin}. Additionally, the structure coefficients
$f^{(B)}_{\ \cdot \ (K)(L)}$ are different from zero which will result in a
qualitatively different dynamics.

\subsection{Isospin particles}

In this case the ansatz (\ref{18})  transforms the first of equations
(\ref{1}) with (\ref{6}) into
\begin{equation}
\frac{\mbox{D}{p^{i}}}{\mbox{D}\tau} = \frac{g I^{(A)} q_{(A)}}{m} F^{i}_{\cdot \ k}
\ p^{k}\ ,
\label{19}
\end{equation}
which has the structure of the equations of motion of a particle
with charge $ e \equiv g I^{(A)} q_{(A)}$ under the influence of the Lorentz force (\ref{2}).
Analogously, one can rewrite  equation (\ref{7}) for the isospin evolution,
\begin{equation}
\frac{\mbox{d}}{\mbox{d}\tau} I^{(A)}= - \Omega \ \epsilon^{(A)}_{\ \cdot \
(B)(C)} q^{(B)} I^{(C)}\ ,\quad
\Omega \equiv \frac{g}{m}A_{i} p^{i}\ .
\label{20}
\end{equation}
Because of the antisymmetry of the Levi-Civita symbols the equations
(\ref{20})
admit a quadratic integral of motion  $I^{(A)} I_{(A)} = {\rm
const} $, which is a  Casimir invariant \cite{Litim}, normalizable to $I^{(A)}
I_{(A)} =1 $.
In addition,  we
obtain from (\ref{20})
\begin{equation}
q_{(A)}\frac{\mbox{d}{I^{(A)}}}{\mbox{d}\tau} \equiv 0\ ,  \quad \rightarrow  \quad
I^{(A)} q_{(A)} \equiv I = {\rm const}\ .
\label{21}
\end{equation}
Using the standard definition
\begin{equation}
\left[ \  \vec{I} \ , \ \vec{\Omega}\ \right]^{(A)} \equiv
\epsilon_{ \ \cdot \ (B)(C)}^{(A)}  I^{(B)}  \Omega^{(C)}\ , \quad
\Omega^{(C)} \equiv \Omega \ q^{(C)}\ ,
\label{22}
\end{equation}
of the vector product, Eq. (\ref{20})  may be written as an equation for the
precession of $\vec{I}$,
\begin{equation}
\frac{\mbox{d}}{\mbox{d}\tau} \ \vec{I} = \left[ \ \vec{I} \ , \ \vec{\Omega}\ \right]\ .
\label{23}
\end{equation}
The ``longitudinal'' component $I^{(A)} q_{(A)}$,
the projection of the dynamical variable  $I^{(A)}$ on the ``rotation axis''
$q^{(A)}$,  remains constant according to (\ref{21}).

\subsection{Colored particles}

Similar to the previous isospin case the color charge evolution equation
(\ref{13})  admits the existence  of a quadratic integral of motion
$Q^{(A)}Q_{(A)} =
{\rm const}$ which is the first Casimir invariant
\cite{Litim}.
The condition (\ref{18}) of parallelism in the
color space provides a second integral of motion $Q^{(A)} q_{(A)}
={\rm const}$, as well. However, the corresponding 8-dimensional evolution equations
\begin{equation}
\frac{\mbox{d}}{\mbox{d}\tau} Q^{(A)}= - \Omega H^{(A)}_{\ \cdot \ (C)} Q^{(C)}\ ,
\quad
\Omega \equiv \frac{g}{m}A_{i} p^{i}\ , \quad
H^{(A)}_{\ \cdot \ (C)} \equiv f^{(A)}_{\ \cdot \ (B)(C)}q^{(B)}\ ,
\label{24}
\end{equation}
are more complicated than the equation (\ref{23}) for the isospin precession.
Different from the $SU(2)$ model there exists
a second Casimir invariant
\begin{equation}
{\cal Q} = d_{(A)(B)(C)} Q^{(A)} Q^{(B)} Q^{(C)} \,,
\label{25}
\end{equation}
where
$d_{(A)(B)(C)}$ are the totally symmetric group coefficients (\ref{17}) of
the given  basic representation of the $SU(3)$ group.
In detail it reads
$$
{\cal Q} = - \frac{1}{\sqrt{3}} (Q^{(8)})^3 + \sqrt{3} Q^{(8)}
\left[(Q^{(1)})^2 + (Q^{(2)})^2 + (Q^{(3)})^2 \right] -
$$
$$
{-} \frac{\sqrt{3}}{2}
Q^{(8)} \left[(Q^{(4)})^2 {+} (Q^{(5)})^2 {+} (Q^{(6)})^2
{+}(Q^{(7)})^2\right]
{+} 3 Q^{(1)} \left[Q^{(4)} Q^{(6)} {+}
Q^{(5)} Q^{(7)} \right] {+}
$$
\begin{equation}
{+} 3 Q^{(2)} \left[{-} Q^{(4)} Q^{(7)} {+} Q^{(5)} Q^{(6)}
\right]
{+} \frac{3}{2} Q^{(3)} \left[(Q^{(4)})^2 {+} (Q^{(5)})^2 {-}
(Q^{(6)})^2 {-} (Q^{(7)})^2\right]\ .
\label{26}
\end{equation}

\section{Gravitational wave background}
\label{GW}

Our aim in this paper is to study the general dynamics outlined so
far in the field of a plane-fronted GW with parallel rays
(PP-wave). We assume the latter to be an exact solution of
Einstein's vacuum field equations with a five-parametric group of
isometries $G_5$, including a covariantly constant null Killing
vector (KV) \cite{ExSol}. Gravitational waves are usually
described either in Fermi coordinates or in the
transverse-traceless (TT)-gauge. For the merits of each of these
choices and for issues of gauge-invariance in the linearized
theory see, e.g., \cite{MTW}. In order to establish a
comprehensive picture we start by sketching our basic setting for
both cases. For computational ease most of the analysis will then
be done in TT-coordinates.

\subsection{PP-wave in Fermi coordinates}

The corresponding line element
\begin{equation}
\mbox{d}s^2 = 2 \mbox{d} \bar{u} \mbox{d}\bar{v}- \mbox{d}y^2
-\mbox{d}z^2 - 2 {\cal H} (\bar{u},y,z) \mbox{d}\bar{u}^2 \ , \quad
\bar{u} = \frac{t - x}{\sqrt{2}}\ , \quad  \bar{v} = \frac{t + x}{\sqrt{2}}
\ ,
\label{27}
\end{equation}
contains a harmonic function ${\cal H}$, obeying
\begin{equation}
\frac{\partial^2 {\cal H}}{\partial y^2} + \frac{\partial^2 {\cal H}}{\partial z^2} = 0 \ ,
\label{28}
\end{equation}
which is quadratic in $y$ and $z$  for a $G_5$ symmetry group. Explicitly, we have \cite{ExSol},
\begin{equation}
2{\cal H}(\bar{u},y,z) = A(\bar{u}) \left( y^2 - z^2 \right) + 2B(\bar{u}) \ y z \ ,
\label{29}
\end{equation}
where $A(\bar{u})$ and
$B(\bar{u})$ are arbitrary functions  of the retarded time $\bar{u}$.
We may define a periodic GW by assuming the variables $A$ and $B$  to be periodic functions of $\bar{u}$.
All non-vanishing components of the Riemann tensor,
$ R_{z \bar{u}z \bar{u}} = - R_{y \bar{u}y \bar{u}} = A(\bar{u})$ and
$R_{y \bar{u}z \bar{u}} = - B(\bar{u})$, are periodic for
this case.
The GW metric in Fermi coordinates is non-singular for arbitrary retarded times
because ${\rm det}(g_{ik}) \equiv - 1 \neq 0$. However, the weak field approximation
${\rm max} |g_{ik}| << 1$  is correct only close to
$ y=0, \ z=0$.

\subsection{PP-wave in TT-gauge}

The line element in TT-gauge has the form
\begin{equation}
\mbox{d}s^{2} = 2\mbox{d}u\mbox{d}v - L^{2} \left[\cosh2\gamma \left(e^{2 \beta}(\mbox{d}x^2)^2 +
e^{-2\beta} (\mbox{d}x^3)^2 \right)+ 2 \sinh{2\gamma}
\mbox{d}x^2 dx^3 \right]\ ,
\label{30}
\end{equation}
where $u=\left(t - x^1 \right)/ \sqrt{2}$ and
$v=\left(t + x^1 \right)/ \sqrt{2}$
are the retarded and the advanced times, respectively.
For this metric the
three KVs which form an Abelian subgroup of
$G_5$ are
\begin{equation}
\xi_{(v)}^i = \delta_v^i, \quad \xi_{(2)}^i = \delta_2^i,\quad \xi_{(3)}^i = \delta_3^i \ .
\label{31}
\end{equation}
The  KV $\xi_{(v)}^i$ is a covariantly constant null vector, orthogonal
to $\xi_{(2)}^i$ and $\xi_{(3)}^i$.
The functions $\beta(u)$ and $\gamma(u)$ are arbitrary.
We shall focus here on the case of periodic functions $\beta(u)$ and $\gamma(u)$. This definition of
periodicity does not, in general, coincide with the definition in Fermi
coordinates given above.
However, both concepts of a periodic GW have the
same weak-field limit,
which in the TT-gauge is characterized by $L=1$,
$|\beta(u)|<<1$ and $|\gamma(u)|<<1$.
Generally, the function $L(u)$  satisfies the Einstein equation \cite{MTW}:
\begin{equation}
\ddot{L} + L \left((\dot{\beta})^2 \cosh^2{2\gamma} + (\dot{\gamma})^2
\right) = 0 \ ,
\label{32}
\end{equation}
where a dot denotes the derivative with respect to
the retarded time $u$.
We assume the hypersurface $u=0$ to be the leading front of the GW with
\begin{equation}
\beta(0) = \gamma(0) = 0\ , \quad L(0)= 1\ , \quad \dot{\beta}(0) =
\dot{\gamma}(0) = \dot{L}(0) = 0\ .
\label{33}
\end{equation}
For the special case $\gamma(u)\equiv 0$
(equivalent to $B \left(\bar{u} \right)\equiv 0$),
corresponding to only one polarization direction, the
transformation relations between Fermi and TT coordinates are
\begin{eqnarray}
\bar{u} &=& u , \quad  \bar{v} = v + \frac{1}{4}
\left[(x^2)^2 (L^2 e^{2\beta})^{\displaystyle \cdot}
+  (x^3)^2(L^2 e^{-2\beta})^{\displaystyle \cdot} \right]\ ,
\nonumber\\
y &=& L e^{\beta} \cdot  x^2 , \quad z = L e^{-\beta} \cdot  x^3 \ ,
\quad
A(u) = \ddot{\beta} + 2 \dot{\beta} \frac{\dot{L}}{L}\ .
\label{34}
\end{eqnarray}
The last formula clarifies the relation between the different periodicity
definitions given above.
For the physically relevant situation where the background factor
$L$ changes only slowly compared with the change of the wave factor
$\beta $  [cf. Ref. \cite{MTW}] the last term in the formula for
$A(u)$ in Eq. (\ref{34}) can be neglected.
Then we have $A(u) = \ddot{\beta}$ and both periodicity definitions coincide.

\section{Particle dynamics: Solutions}
\label{solution}

We are interested here in the particle dynamics in given external
gravitational and Yang-Mills fields. The restriction (\ref{18}) to
Yang-Mills fields with ``parallel potentials'' simplifies the
Kerner-Wong equations since the quantities $I^{(A)}q_{(A)}$ and
 $Q^{(A)}q_{(A)}$ remain constant.
 We introduce the cumulative
symbol $\sigma$ for either $e$, or $gI^{(A)}q_{(A)}$, or $gQ^{(A)}q_{(A)}$,
which allows us to write the equation of motion with either (\ref{2}), or
(\ref{6}), or (\ref{10}) in the unified form
\begin{equation}
\frac{\mbox{D}p^i}{\mbox{D}\tau} =
\frac{\sigma}{m} F^{i}_{\cdot \ k} \ p^{k} ,
\quad \frac{\mbox{d} x^i}{\mbox{d}\tau} = \frac{p^i}{m}\ .
\label{35}
\end{equation}
The orthogonality of the force to the particle momentum corresponds to the quadratic integral
\begin{equation}
g^{ik} p_i p_k = m^2\ .
\label{36}
\end{equation}
One may solve this relation for one of the components of the momentum.
In Fermi coordinates we have
\begin{equation}
p _{\bar{u}} = \frac{1}{2 p _{\bar{v}}}
\left[ m^2  + p^2_y + p^2_z - 2 {\cal H}(\bar{u},y,z) \ p^2 _{\bar{v}}\right]\
\label{37}
\end{equation}
The analogous formula in TT coordinates  is
\begin{equation}
p _{u}
=
\frac{1}{2 p _{v}}
\left[m^{2} - g^{\alpha \beta}\left(u \right)p _{\alpha}p _{\beta }
\right]\ ,
\label{38}
\end{equation}
where greek indices run from 2 to 3.
Both in Fermi and in TT coordinates the covariantly constant null KV
has the form $\xi^i_{\left(\bar{v}\right)} = \delta^i_{\bar{v}}$ and
$\xi^i_{\left(v \right)} = \delta^i_v$, respectively.
Below we shall restrict ourselves to fields which satisfy
\begin{equation}
\xi _{\left(\bar{v} \right)}^{i}F _{ik} = \xi _{\left(v \right)}^{i}F _{ik}
= 0 \ .
\label{39}
\end{equation}
This implies that the quantities
$\xi^i_{(\bar{v})} p_i$ and $\xi^i_{(v)} p_i$ are integrals of motion
(see e.g. \cite{MTW}),
\begin{equation}
\xi^i_{(\bar{v})} p_i = \xi^i_{(v)} p_i = p_v = C_v = {\rm const} \ .
\label{40}
\end{equation}
Using the general relationship
\begin{equation}
m\frac{d u}{d\tau} = p^u = p ^{\bar{u}} = p_v = C_v \ ,
\label{41}
\end{equation}
one can reparametrize the remaining equations
for $C_v \neq 0$, by means of the
linear formula (notice that $\bar{u}=u$ [cf. Eq. (\ref{34})] holds also in
the general case)
\begin{equation}
\tau = \tau_0 + \frac{m}{C_v} u \ .
\label{42}
\end{equation}
We start our solution procedure by first recalling the reference case of neutral particles.

\subsection{Neutral particles}
\label{neutral}

For a vanishing generalized charge $\sigma $  the equations of motion in TT coordinates
are immediately integrated.
The result is
\begin{equation}
p_v(u) = C_v\ , \  p_2(u) = C_2\ , \  p_3(u) = C_3 \ ,\
p_u = \frac{1}{2C_v} \left[ m^2  {-} g^{\alpha \beta}(u) \ C_\alpha C_\beta
\right]  ,
\label{43}
\end{equation}
where $C _{2}$ and $C _{3}$ are constants. A particle moving in
direction of the GW propagation before the infall of the latter,
i.e., $p _{2}=p _{3}=0$ will not change its direction. If we
additionally have $C_v = \frac{mc}{\sqrt{2}}$, the particle is at
rest both before and after the GW infall, since $p_1(u) = p_2(u) =
p_3(u) = 0$ and $p_0 = mc$. An observer at rest, characterized by
a four-velocity $V^i = \frac{1}{\sqrt{2}} (\delta^i_u +
\delta^i_v) \equiv \delta^i_0 $, would measure the (invariant)
particle energy ${\cal E} \equiv p^k V_k = \frac{1}{\sqrt{2}} (p_u
+ p_v)$. For neutral particles $p_v$ and $p_u$ are given in
Eq.~(\ref{43}). Corresponding expressions for charged particles
will be obtained below.

In Fermi coordinates the situation is as follows. The system
(\ref{35}) reduces to the set of equations
$$
\dot{y} = - C^{-1}_v p_y \ , \quad
\dot{z} = - C^{-1}_v p_z \ , \quad
$$
\begin{equation}
C^{-1}_v \dot{p}_y = -A(u) \ y - B(u) \ z \ , \quad
C^{-1}_v \dot{p}_z = A(u) \ z - B(u) \ y \ ,
\label{44}
\end{equation}
for $y$, $z$, $p_y$, and $p_z$. If the latter set of quantities is known, the component
$p _{u}$ follows via (\ref{37}).
The quantity $\bar{v}(u)$ may be found by solving the equation
\begin{equation}
\dot{\bar{v}} =  C^{-1}_v p_u + 2 {\cal H}(u,y,z)\ .
\label{45}
\end{equation}
The set (\ref{44}) is a linear, homogeneous first-order system  of
differential equations.
For a GW field
with  polarization $B(u) \equiv 0$ and with $A$ depending on the retarded time
via the dimensionless variable $ku$, it can be written as
\begin{equation}
y ^{\prime \prime } - {\cal A}(ku) \ y = 0\ , \quad
z^{\prime \prime } + {\cal A}(ku) \ z = 0\ .
\label{46}
\end{equation}
Differentiation with respect to $ku$ is denoted by a prime and
${\cal A}\left(ku \right)\equiv  A \left(u \right)/k ^{2}$.
For $A(u) \equiv 0$, i.e. in the absence of a gravitational field, $y(u)$ and $z(u)$ are linear
functions of the retarded time (and of the affine parameter
$\tau$), and the particle has  constant momentum.
For a periodic function $A(u)$ the equations (\ref{46})  are of the type of
Hill's  equation \cite{Stoker,McLa,YaSta}.
The solutions are Hill functions.
For a dependence
${\cal A}\left(ku\right)=\delta + \epsilon \cos (ku)$ where $\delta $ and $\epsilon $
are constants, Eqs. (\ref{46}) reduce to Mathieu equations
which have solutions of the type \cite{Stoker,Arscott,YaSta}
\begin{equation}
y \propto  \exp{\left[\mu ku \right]}\phi \left(ku \right)
+ \exp{\left[-\mu ku \right]}\psi \left(ku \right)
\,,
\label{47}
\end{equation}
where $\phi $ and $\psi $  are periodic functions with the period of
${\cal A} \left(ku \right)$,  i.e. in the present case, $\phi \left(ku + 2\pi
\right)=\phi \left(ku \right)$  and
 $\psi \left(ku + 2\pi \right)=\psi
\left(ku \right)$. The solutions consist of products  of an exponential
function and a periodic function of period $2\pi$.
The characteristic
exponent $\mu $ is a complex constant.
For $Re \left(\mu  \right) =0$  the
solution is stable.
In general, it is not periodic again but it is
oscillating (\cite{McLa}, p.115).
For $Re \left(\mu  \right)\neq 0$ either
the first or the second exponential function in (\ref{47})  is unbounded and
the solution is unstable.
The stability properties of Mathieu's equations are
well known in the literature and may  be visualized by stability regions in an
$\delta $ - $\epsilon $ diagram (see \cite{Stoker}, Eq.(4.1) and Fig. 5.1).
Let's consider the stability region which is closest to the origin
$\delta =\epsilon =0$. For small positive values of $\delta $ and $\epsilon $
the boundary of this region is determined by the line
$\delta = \frac{1}{4} - \frac{1}{2}\epsilon $.
Applied to the case $A \left(u\right)=A _{0}\cos (ku)$, i.e. $\delta =0$ and
$\epsilon =A _{0}/k ^{2}$, this means stable solutions for
$\epsilon =\frac{A_0}{k^2}< 1/2$.
Since $A _{0}= \ddot{\beta }\left(0 \right)= \beta _{0}k ^{2}$, we have also
$\beta _{0}< 1/2$.
Under this condition neutral particles are parametrically oscillating in
Fermi coordinates.

While equations of the type of Mathieu's equation and questions of
stability will play an essential role in the following
investigations of the dynamics of charged particles (see
subsection \ref{sandwich} below), it is obvious that the
description for neutral particles is more involved in Fermi
coordinates. Therefore, for computational ease and  in order to
separate charge effects from the neutral particle motion we shall
perform the following analysis in TT coordinates.

\subsection{Charged particles}

To obtain exactly solvable models
for the particle motion  we resort to simple field configurations
$F^{(A)}_{ik}$.
For particles with electric charge we focus on the motion of the latter in
a constant homogeneous magnetic field $H_0$ orthogonal to the GW front plane,
which corresponds to a Maxwell tensor
\begin{equation}
F_{jk} = H_0  \left(\delta^{2}_{j} \delta^{3}_{k} -
\delta^{3}_{j}\delta^{2}_{k} \right)\ .
\label{48}
\end{equation}
A corresponding generalization for non-Abelian fields with
parallel potentials according to (\ref{18}) is
\begin{equation}
F^{(A)}_{jk} = q^{(A)} M  \left(\delta^{2}_{j} \delta^{3}_{k} -
\delta^{3}_{j}\delta^{2}_{k} \right), \quad M = {\rm const}\ .
\label{49}
\end{equation}
Both (\ref{48}) and (\ref{49}) satisfy (\ref{39}) with (\ref{31}).
This constitutes a model in which the gravitational wave and the
fields (\ref{48}) or (\ref{49}) are given, external fields which
are independent of each other. It may provide the basis of a
perturbative treatment with respect to the GW amplitude within a
linearized theory.  It is worth mentioning, that the expressions
(\ref{48}) and (\ref{49}) are also solutions of the Maxwell- and
Yang-Mills equations, respectively, on the background of the exact
GW (\ref{30}) (or (\ref{27})). This allows a study of the
corresponding field dynamics on a GW background, which, however,
is not the purpose of the present paper.

For the equations of motion in TT coordinates we obtain
\begin{equation}
\frac{\mbox{d}p_{2}}{\mbox{d}\tau}=\frac{M \sigma}{m}
\left(g^{32} p_{2} + g^{33} p_{3} \right)\ ,
\quad
\frac{\mbox{d}p_{3}}{\mbox{d}\tau}= - \frac{M \sigma}{m} \left(g^{22} p_{2} +
g^{23} p_{3} \right)\ .
\label{50}
\end{equation}
Equivalent second-order equations are
\begin{equation}
\frac{\mbox{d}^2p_{2}}{\mbox{d}u^2}
+ R_{2}(u)\frac{\mbox{d}p_{2}}{\mbox{d}u} + W_{2}(u) p_{2}=0 \ , \quad
p_3 = \frac{1}{g^{33}} \left(\frac{1}{\Pi} \dot{p}_2 - g^{23} p_2 \right) \ ,
\label{51}
\end{equation}
or
\begin{equation}
\frac{\mbox{d}^2p_{3}}{\mbox{d}u^2}
+ R_{3}(u)\frac{\mbox{d}p_{3}}{\mbox{d}u} + W_{3}(u) p_{3}=0 \ , \quad
p_2 = \frac{1}{g^{22}} \left(- \frac{1}{\Pi} \dot{p}_3 - g^{23} p_3 \right)\ ,
\label{52}
\end{equation}
where we have introduced the notations
\begin{eqnarray}
R_{2}(u)&=&-\frac{\dot g^{33}(u)}{g^{33}(u)} =2\left(\frac{\dot{L}}{L} -
\dot{\beta} -\dot{\gamma} \tanh(2\gamma)\right) \ ,\nonumber\\
R_{3}(u)&=&-\frac{\dot{g}^{22}(u)}{g^{22}(u)}=
2\left(\frac{\dot{L}}{L} + \dot{\beta} -\dot{\gamma} \tanh(2\gamma)\right) \ ,
\label{53}
\end{eqnarray}
and
\begin{eqnarray}
W_2(u)&=& \frac{\Pi^2}{L^4} + \frac{2\Pi}{L^2}
\left(\dot{\beta}\sinh(2\gamma) - \frac{\dot{\gamma}}{\cosh(2\gamma)}
\right)\ ,\nonumber\\
W_3(u)&=& \frac{\Pi^2}{L^4} + \frac{2\Pi}{L^2} \left(
\dot{\beta}\sinh(2\gamma) + \frac{\dot{\gamma}}{\cosh(2\gamma)} \right) \ ,
\label{54}
\end{eqnarray}
with
\begin{equation}
\Pi=\frac{M \sigma}{C_{v}} = {\rm const}\ .
\label{55}
\end{equation}
The substitution $\beta \to  - \beta$ converts
$R_3$ into $R_2$ and vice versa, while $W_3$ is
obtained from $W_2$ by $\beta \to  - \beta$ and
simultaneously $\Pi \to - \Pi$.

By the substitution
\begin{equation}
p_{\alpha}= Z_{\alpha}(u)
\exp\left\{-\frac{1}{2}{\int_{0}^{u}R_{\alpha}(\zeta)d\zeta}\right\}\
\label{56}
\end{equation}
the equations (\ref{51}) and (\ref{52}) may be transformed into the
Hill equations
\begin{equation}
\ddot{Z}_{\alpha} + F_{\alpha}(u) Z_{\alpha}=0\ ,
\label{57}
\end{equation}
where
\begin{equation}
F_{\alpha}=W_{\alpha}-\frac{R^2_{\alpha}}{4}-
\frac{\dot{R}_{\alpha}}{2}\ .
\label{58}
\end{equation}
The detailed form of the relations (\ref{56}) and (\ref{58}) is
\begin{equation}
p_{2}{=} Z_{2}(u)\sqrt{\cosh(2\gamma)} \ \frac{e^{\beta}}{L}\ , \
p_{3}{=} Z_{3}(u) \sqrt{\cosh(2\gamma)} \ \frac{e^{-\beta}}{L}\ , \
p_{\alpha}(0) {=} Z_{\alpha}(0) \equiv C_{\alpha} ,
\label{59}
\end{equation}
and
$$
F_{2}(u) = \frac{\Pi^2}{L^4} + \frac{2\Pi}{L^2} \left(
\dot{\beta}\sinh(2\gamma) - \frac{\dot{\gamma}}{\cosh(2\gamma)} \right) +
\ddot{\beta} + \ddot{\gamma} \tanh(2\gamma)
$$
\begin{equation}
+ \frac{3(\dot{\gamma})^2}{(\cosh(2\gamma))^2}
+(\dot{\beta})^2(\sinh(2\gamma))^2 + 2\frac{\dot{L}}{L}\dot{\beta} +
2\frac{\dot{L}}{L} \dot{\gamma}\tanh(2\gamma) - 2 \dot{\gamma}\dot{\beta}
\tanh(2\gamma) \ ,
\label{60}
\end{equation}
$$
F_{3}(u) = \frac{\Pi^2}{L^4} + \frac{2\Pi}{L^2} \left(
\dot{\beta}\sinh(2\gamma) + \frac{\dot\gamma}{\cosh(2\gamma)} \right)
-  \ddot{\beta} + \ddot{\gamma}\tanh(2\gamma)
$$
\begin{equation}
+
\frac{3(\dot{\gamma})^2}{(\cosh(2\gamma))^2} +
(\dot{\beta})^2(\sinh(2\gamma))^2 - 2\frac{\dot{L}}{L}\dot{\beta} +
2\frac{\dot{L}}{L} \dot{\gamma}\tanh(2\gamma) + 2 \dot{\gamma}\dot{\beta}
\tanh(2\gamma) \ ,
\label{61}
\end{equation}
respectively.
In a next step we have to solve Hill's equations (\ref{57}).

\subsubsection{General solution of Hill's equation}

The structure of the solutions of the linear, second-order
differential equations (\ref{57}) is
\cite{Stoker,Arscott,McLa,YaSta,Meix}
\begin{equation}
Z_2(u) = C_2 \  H_2(u) - \Pi \ C_3 \ H_3(u) \ .
\label{62}
\end{equation}
The functions $H_{\alpha}(u)$ satisfy the initial
conditions
\begin{equation}
H_2(0) = 1, \quad \dot{H}_2(0) = 0, \quad
H_3(0) = 0, \quad \dot{H}_3(0) = 1\ ,
\label{63}
\end{equation}
and represent the fundamental solutions of Hill's equation
(\ref{57}) with unitary Wronsky determinant.
For $Z_3(u)$ we have
\begin{equation}
Z_3(u) =  C_3 \ H^*_3 + C_2 \Pi \ H^*_2 \ ,
\label{64}
\end{equation}
where
\begin{equation}
H^*_2 = - \frac{L^2}{\Pi^2 \cosh{2\gamma}} \left[\dot{H}_2 +
H_2 \left(\dot{\gamma}\tanh{2\gamma} + \dot{\beta} - \frac{\dot{L}}{L} -
\frac{\Pi }{L^2} \sinh{2\gamma}\right) \right] \,,
\label{65}
\end{equation}
\begin{equation}
H^*_3 = \frac{L^2}{\cosh{2\gamma}} \left[\dot{H}_3 +
H_3 \left(\dot{\gamma}\tanh{2\gamma} + \dot{\beta} - \frac{\dot{L}}{L} -
\frac{\Pi }{L^2} \sinh{2\gamma}\right) \right] \,,
\label{66}
\end{equation}
\begin{equation}
H^*_2(0)= \dot{H}_2(0)= 0 \,, \quad
H^*_3(0)= \dot{H}_3(0)= 1 \,.
\label{67}
\end{equation}
In the absence of gravitational radiation, i.e. for
$\beta=\gamma \equiv 0, \ L \equiv 1$,
the functions $F_\alpha$ in equation (\ref{58}) reduce to
\begin{equation}
F_2(u)  =  F_3(u) = {\rm const} = \Pi^2 \,.
\label{68}
\end{equation}
Equation (\ref{57}) then describes harmonic oscillations with
\begin{equation}
H_2 = H^*_3 \equiv \cos{\Pi u}, \quad H_3 = H^*_2
\equiv \frac{1}{\Pi} \sin{\Pi u}\ .
\label{69}
\end{equation}
Since with (\ref{42}) and (\ref{55}) we have $\Pi u \to \Omega_H
\tau $ where $\Omega_H \equiv \frac{e H_0}{mc}$ is the Larmor
frequency, we recover the corresponding particle rotation in flat
spacetime. Generally, the functions $H_2$, $H_3$, $H^*_2$ and
$H^*_3$ cannot be written in terms of elementary functions but are
given as series representations. In the following subsection we
shall be interested in expressions for $F_2$ and $F_3$ for which
the Hill equations (\ref{57}) specify to Mathieu equations. Then
the functions $H_\alpha$ can be expanded in powers of the GW
amplitude, where the zeroth order is given by (\ref{69}).

The analysis so far may be summarized by writing the solution of the equations of motion
(\ref{51}) and (\ref{52}) in the compact and elegant matrix form
\begin{equation}
\left(\begin{array}{c}p_2 \\p_3\end{array}\right) =
{\bf H}(u) \cdot
\left(\begin{array}{c}C_2 \\C_3\end{array}\right)\ ,
\label{70}
\end{equation}
where
\begin{equation}
{\bf H}(u) \equiv
\frac{\sqrt{\cosh2\gamma}}{L}
\left(\begin{array}{cc} e^{\beta} &  0 \\ 0 &
e^{-\beta} \end{array} \right) \cdot
\left(\begin{array}{cc}
H_2(u)  & - \Pi H_3(u)  \\
\Pi H_2^{*}(u)  & H_3^{*}(u)
\end{array}\right)\ .
\label{71}
\end{equation}
The set of equations (\ref{70}) and (\ref{71}) represents the general solution for the momentum of the
charged particle in the GW field (\ref{30}) and the Yang-Mills field
(\ref{49}).

\subsubsection{A simple model}
\label{sandwich}

Now we apply the general formalism to a ``sandwich'' GW (see, e.g., \cite{MTW})
with polarization
$\gamma =0$.  Let $\beta $ be different from zero during a
finite retarded time interval $T$,  i.e., $\beta = 0$  for $u \leq 0$ and
$u \geq T$.
Within the interval $0 < u < T$ we assume $\beta $ to be periodic according to
\begin{equation}
\beta(u)=\beta_{0}(1-\cos(ku))\ , \quad \quad
\left(0 < u < T \right)\ .
\label{72}
\end{equation}
This implies $\beta(0)=\beta(2\pi/k)=0$ and $\dot\beta(0)=\dot\beta(2\pi/k)=0$.
Furthermore, the time scale $T$ is assumed to be small compared with the scale on
which the background factor $L$ changes \cite{MTW}.
Under these conditions we may neglect the $\dot{L}/L$ terms  in (\ref{60}) and
(\ref{61}) and use the latter expressions with $L=1$.
For this situation the potentials $F _{2}$ and $F _{3}$  reduce to
\begin{eqnarray}
F_2 &=& \Pi^2 + \beta _{0}k ^{2}\cos \left(ku \right)
= \Pi^2 - R^2_{\cdot u2u}\ ,\nonumber\\
F_3 &=& \Pi^2 - \beta _{0}k ^{2} \cos \left(ku \right) =  \Pi^2 -
R^3_{\cdot u3u}
\label{73}
\end{eqnarray}
in the interval $0<u<T$.
Replacing now the variable $u$ by $ku$  and
denoting the derivative with respect to $ku$ again by a prime,
the equations (\ref{57}) with
(\ref{73}) specify to
\begin{eqnarray}
Z _{2}^{\prime \prime }
+ \left(\frac{\Pi ^{2}}{k ^{2}} + \beta _{0}\cos \left(ku \right)\right)
Z _{2} &=&0     \ ,    \nonumber\\
Z _{3}^{\prime \prime }
+ \left(\frac{\Pi ^{2}}{k ^{2}} - \beta _{0}\cos \left(ku \right)\right)
Z _{3}&=&0 \ .
\label{74}
\end{eqnarray}
Both $Z _{2}$ and $Z _{3}$ obey Mathieu equations.
In the absence of the GW, i.e. for $\beta \equiv 0$, we have $F_{\alpha}(ku) = \Pi^2/k ^{2}={\rm const}$
(here we have used the redefinition
$F _{\alpha}\left(ku \right)\equiv  F _{\alpha}\left(u \right)/k ^{2}$ )
and the equations of motion  reduce to harmonic oscillator equations with
solutions
\begin{equation}
Z _{2} =p _{2} = C _{2}\cos \left(\Pi u \right)
- C _{3}\sin \left(\Pi u \right)
\label{75}
\end{equation}
and
\begin{equation}
Z _{3} = p _{3} = C _{3}\cos \left(\Pi u \right) + C _{2}\sin \left(\Pi u \right)\ .
\label{76}
\end{equation}
Replacing here $\Pi $ and $u$  according to (\ref{55})  and (\ref{42}), we
find a particle  rotation in the  $x ^{2} 0 x ^{3}$ plane
with the angular velocity $\Omega_M \equiv \frac{M \sigma}{m}$, which
is, of course, the analogue of the Larmor frequency.
Immediately after the wave front, i.e. at $u=0_{+0}$, we have
\begin{equation}
F_2(0) = \frac{\Pi ^{2}}{k ^{2}}+ \beta _{0}\ ,
\quad
F_3(0) = \frac{\Pi ^{2}}{k ^{2}}- \beta _{0}\ .
\label{77}
\end{equation}
The jump $\beta_0 k^2$ of the curvature tensor at the front makes
the evolutions of the $p_2$ and $p_3$ components different. They
begin to oscillate  with different frequencies and the particle
trajectory is no longer circular. Corresponding features hold for
the second polarization $\beta \equiv 0$ and $\gamma \neq 0$, for
which $F _{2}$  and $F _{3}$ differ in the term linear in $\Pi$
[cf. Eqs. (\ref{60}) and (\ref{61})].

The general solutions of Eq.~(\ref{74}) are of the type of
``cosine elliptic'' and  ``sine elliptic'' functions (see
\cite{McLa,Meix}). As already mentioned, the latter may be
expanded in powers of $\beta_0$ with the zeroth-order terms
(\ref{75})  and (\ref{76}).

Eqs.~(\ref{74}) are of the same type as Eqs.~(\ref{46}).
With the identifications $\delta \to \Pi^2 /k^2$ and $\epsilon \to \beta_0$,
the stability discussion  mentioned in subsection
\ref{neutral} may be applied here as well (see \cite{Stoker}, Eq.(4.1) and
Fig. 5.1).
Depending on the
parameter combinations  the  solutions may be stable or unstable.
Within the stable regions the functions $Z _{2}$ and $Z _{3}$  are
parametrically oscillating
which, according to (\ref{59}) implies a corresponding behavior of the particle
momenta.
The regions of stable solutions are characterized by stability
zones in a
$\left(\Pi /k \right)^{2} \times \beta _{0}$ plane which are
connected together
at the points $\left(\Pi /k \right)^{2}=n ^{2}/4$, $\beta_{0}= 0$, where $n$
is an integer \cite{Stoker}.

The analysis of the neutral particle motion in subsection \ref{neutral}
corresponds to the case
$\Pi = n = 0$. For $\Pi \neq 0$ the relevant values for the transition points
are $n=1, 2, ....$. For $n=1$ we have
$\Pi/k = 1/2$. These points on the
axis $\beta _{0}= 0$ (which corresponds to
the absence of the GW) are the
only transition points between stable regions
which also belong to the stable
region. All other boundary points of the
stable regions are unstable
points.
Consequently, any deviation from $\beta_{0}=0$,
i.e. even a GW with arbitrary weak amplitude $\beta _{0}$, induces an
instability  in these critical points. In particular, this is true for the point
$\Pi/k = 1/2$ (see \cite{Stoker}, Fig. 5.1).
This demonstrates that
parametric instabilities are a generic phenomenon for the motion of particles
in all the cases  considered here.
While we have obtained this result in TT
coordinates, the transformations (\ref{34}) allow us to find the corresponding
particle momenta in Fermi coordinates as well.
The relevant transformations are
$$
p_{\bar{v}} = p_v = C_v \,,
\quad
p_y = p_2(u) \frac{e^{-\beta}}{L} - C_v \left( \frac{\dot{L}}{L} +
\dot{\beta}\right) y(u) \,,
$$
\begin{equation}
p_z = p_3(u) \frac{e^{\beta}}{L} - C_v \left( \frac{\dot{L}}{L} -
\dot{\beta}\right) z(u) \,,
\label{78}
\end{equation}
where
$$
y(u) = L e^{\beta} \left[ y(0) - \int_0^u d \zeta p_2(\zeta) L^{-2}(\zeta)
e^{-2\beta(\zeta)} \right]       \ ,
$$
\begin{equation}
z(u) = L e^{-\beta} \left[ z(0) - \int_0^u d \zeta p_3(\zeta) L^{-2}(\zeta)
e^{2\beta(\zeta)} \right] \ .
\label{79}
\end{equation}
It is interesting to realize that there are astrophysical
situation for which the existence  of such kind  of instabilities
might be relevant. This can be seen with the help of the following
order-of-magnitude estimates. Eq.~(\ref{55}) may be written as
$\Pi =\Omega_{M}m/C_{v}$ with $\Omega_{M}= M \sigma /m$. For the
electromagnetic case one has $\sigma = e$  and $M=H_{0}$. The
interstellar magnetic field is of the order $3\div6 \cdot 10^{-6}
{\rm Oe}$ \cite{Kaplan}. The integral of motion $C_{v}$ is equal
to $C_v =(\sqrt{m^2+\vec{p}^2(0)} - p^1(0))/\sqrt{2}$. For
nonrelativistic particles $C_v \propto m/\sqrt{2}$ and $\Pi
\propto \omega_H \sqrt{2}$. (The coefficient $\sqrt{2}$ disappears
if we use the natural parameter $\tau$ instead of retarded time
$u$). Using the estimate \cite{Kaplan} $\omega_H \propto  10^7
\cdot H_0 ({\rm Hz})$ we find $\omega_H\approx 10^1 \div 10^2$ Hz.
This is well within the typical range $1-10^3 {\rm Hz}$ for the
frequency $k$ of a GW, generated by rapidly rotating neutron stars
(pulsars). Thus, for non-relativistic particles $\Pi$ may be  of
the same order as the GW frequency $k$. The situation is different
for ultrarelativistic particles. Since for particles that move in
propagation direction of the GW ($p^1(0)>0$), we have $C_v \to 0$
for $m \to 0$. Consequently, the quantity $\Pi$ becomes very large
[cf. Eq.~(\ref{55})], such that $\Pi >> k$. However, if the
particles move in the opposite direction, i.e. $p^1(0)<0$, any
value of $C_v$ is possible. In particular, $\Pi$ is not excluded
to be in the range of about $10^{-3} {\rm Hz}$ which is typical
for infra-low-frequency GW from relativistic compact binaries.

Since an ensemble of particles will generally not be characterized by a single
value of $C _{v}$ but by a distribution, the quantity $C _{v}$  may play  the
role of a tuning parameter in the following sense. Let us associate a mean
value $\langle C _{v}\rangle$ for the system as a whole and let the system be
outside the resonance $\omega_H /k = n / 2$ for this
$\langle C_{v}\rangle$, but not very far from it.
Since $\Pi =\omega_{H}m/C_{v}$, we will very likely
find a particular particle with a specific $C _{v}$ such that for this
particle $\Pi/k = n/2$ exactly.
Consequently, certain particles of the
ensemble might be resonant under these conditions.

\section{Evolution of the spin four-vector}
\label{spin}

In this section we focus on the solution of Eq.~(\ref{5}) in external gravitational and magnetic fields.
The equations (\ref{4}) and (\ref{5}) admit two
integrals of motion, namely $p_i S^i = 0$ and  $S_i S^i = -
E_0^2 = {\rm const}$ \cite{BMT}.
In the present context this amounts to
\begin{equation}
p_v S_u + p_u S_v + p_\alpha S^\alpha = 0, \quad
2 S_u S_v + S_\alpha S^\alpha = - E_0^2 \ ,
\label{80}
\end{equation}
which may be used to eliminate the components
$S_v$ and $S_u$ according to
\begin{equation}
S_u  = \frac{1}{2p_v} \left[ - p_\alpha S^\alpha  \mp \sqrt{(p_\alpha
S^\alpha)^2 + (m^2 - p_\alpha p^\alpha)(E^2_0 + S_\alpha S^\alpha)}
\right]\ ,
\label{81}
\end{equation}
and
\begin{equation}
S_v  = \frac{1}{2p_u} \left[ - p_\alpha S^\alpha  \pm \sqrt{(p_\alpha
S^\alpha)^2 + (m^2 - p_\alpha p^\alpha)(E^2_0 + S_\alpha S^\alpha)}
\right]\ ,
\label{82}
\end{equation}
respectively. After the reparametrization (\ref{42})
the remaining equations are
\begin{equation}
\frac {d S_\alpha}{d u} =
\frac{1}{2} g^{\sigma\rho}\dot{g}_{\rho\alpha}
\left(S_\sigma  - p_\sigma \frac{S_v}{C_v} \right)
+ \frac{e}{2 C_v} \left[ \ g F_{\alpha k} \
S^{k} + \frac{(g - 2)}{m^2}  p_\alpha F_{kl} S^k p^l \right] \ .
\label{83}
\end{equation}
Furthermore, the equation for  $S_v$ is
\begin{equation}
\frac {d S_v}{d u} =
\frac{e(g - 2)}{2m^2} F_{kl} S^k p^l .
\label{84}
\end{equation}
In the
following we shall restrict ourselves to the exactly integrable case $g=2$.
Under this condition  we find from (\ref{84}) that
\begin{equation}
S_v = const = E_v \ .
\label{85}
\end{equation}
After the substitution
\begin{equation}
S_\alpha = X_\alpha + \frac{E_v}{C_v} p_\alpha \ ,
\label{86}
\end{equation}
where $p_\alpha$ is assumed to be given by Eq.~(\ref{70}) in terms
of Hill (or Mathieu) functions, we obtain a homogeneous equation
for the new variable $X_\alpha$,
\begin{equation}
\frac {\mbox{d} X_\alpha}{\mbox{d}u} =
\frac{1}{2} g^{\sigma\rho}\dot{g}_{\rho\alpha} X_\sigma
+ \frac{e}{C_v}  F_{\alpha \sigma} g^{\sigma \lambda}\ X_{\lambda}\ .
\label{87}
\end{equation}
It is convenient to write this equation in the
matrix form
\begin{equation}
\frac{\mbox{d}}{\mbox{d}u}{\bf X} = \left( {\bf A} + \frac{\Pi}{L^2} \ {\bf B} \right)
\cdot  {\bf X}\ ,
\label{88}
\end{equation}
where ${\bf X}$ is a column vector with elements  $X_2$ and $X_3$.
The two-dimensional matrices ${\bf A}$ and ${\bf B}$ have the structures
\begin{equation}
{\bf A} {=}
\left(\begin{array}{cc}
\frac{\dot{L}}{L}+\cosh ^2(2\gamma )\dot{\beta } & e^{2\beta}(\dot{\gamma}-
\sinh(2\gamma)\cosh(2\gamma)\dot{\beta }) \\
e^{-2\beta }(\dot{\gamma}+\sinh (2\gamma )\cosh(2\gamma
)\dot{\beta }) & \frac{\dot{L}}{L}-\cosh ^2(2\gamma )\dot{\beta }
\end{array}\right)\ ,
\label{89}
\end{equation}
and
\begin{equation}
{\bf B} \equiv
\left(\begin{array}{cc}
\sinh2\gamma  & - e^{2\beta}\cosh 2\gamma  \\
e^{- 2\beta}\cosh 2\gamma  & - \sinh 2\gamma
\end{array}\right)\ ,
\label{90}
\end{equation}
respectively.
Equations (\ref{50}) may be written  in matrix form as well:
\begin{equation}
\left(\begin{array}{c}p_2 \\p_3\end{array}\right)^{\displaystyle\cdot}
= \frac{\Pi}{L^2} \ {\bf B} \cdot
\left(\begin{array}{c}p_2 \\p_3\end{array}\right)\ .
\quad
\label{91}
\end{equation}
For the derivative of the combination  $p_2^2 + p_3^2$ we obtain
\begin{equation}
\frac{\mbox{d}}{\mbox{d} u} \left( p_2^2 + p_3^2 \right) = \frac{2 \Pi}{L^2}
\left[ \left(p_2^2 - p_3^2 \right) \sinh2\gamma - 2 p_2 p_{3}
\cosh2\gamma \sinh2\beta \right]\ .
\label{92}
\end{equation}
In general, the right-hand side of this equation is different from zero,
i.e., the particle motion is no longer circular in the GW field.

Eq.~(\ref{88}) may be further simplified by changing to a new variable ${\bf Y}$, defined by
\begin{equation}
{\bf X} =  {\bf T} \cdot {\bf Y}\ ,
\label{93}
\end{equation}
where
${\bf T}$ is supposed to satisfy the differential
equation
\begin{equation}
\frac{\mbox{d}}{\mbox{d}u}{\bf T} = {\bf A} \cdot {\bf T}\ .
\label{94}
\end{equation}
This procedure [cf. \cite{B1,BKZ}] removes the ${\bf A}$ term in Eq.~(\ref{88})and gives rise to
the equation
\begin{equation}
\frac{\mbox{d}}{\mbox{d}u}{\bf Y} =  \frac{\Pi}{L^2} \
\hat{\bf B} \cdot {\bf Y}
\label{95}
\end{equation}
for ${\bf Y}$
with
\begin{equation}
\hat{\bf B} \equiv
{\bf T}^{-1} \cdot {\bf B} \cdot {\bf T} \ .
\label{96}
\end{equation}
By direct calculation one checks that the
matrix
\begin{equation}
{\bf T} =
L \cdot \left(\begin{array}{cc}e^{\beta } & 0 \\ 0 & e^{-
\beta}\end{array}\right) \cdot
\left(\begin{array}{cc}
\cosh \gamma  & \sinh \gamma  \\
\sinh \gamma  & \cosh \gamma
\end{array}\right) \cdot
\left(\begin{array}{cc}
\cos \psi  & -\sin \psi  \\
\sin \psi  & \cos \psi
\end{array} \right)\ ,
\label{97}
\end{equation}
where
\begin{equation}
\psi \equiv \int\limits_0^u \dot{\beta} \sinh 2\gamma \mbox{d}u\  ,
\label{98}
\end{equation}
satisfies the equation (\ref{94}).
The determinants of each of the
three two-dimensional matrices in (\ref{97})
are equal to one. In the absence of the GW field all of them
are identical to ${\bf I}$, i.e.,
\begin{equation}
{\bf T}(0) =  {\bf I} \equiv
\left(\begin{array}{cc}
1  &  0  \\
0  &  1
\end{array} \right)\ .
\label{(99}
\end{equation}
The structure of the third matrix on the right-hand side of Eq.~(\ref{97})
suggests
the interpretation as a gravitationally induced rotation  with phase
$\psi(u)$ and frequency
$\dot{\psi}(u)$.
For either of the polarizations $\gamma  =0$ or $\beta =0$, however, we
have $\psi = 0$
and the third matrix reduces to ${\bf I}$.

Direct calculation of the matrix $ \hat{\bf B}$ in (\ref{96}) with the help of
the expressions (\ref{90}), (\ref{97}) and (\ref{98}) yields the
surprisingly simple result
\begin{equation}
\hat{\bf B} \equiv {\bf T}^{-1} \cdot {\bf B} \cdot
{\bf T} = \left(\begin{array}{cc}
0  & - 1  \\
1  &  0
\end{array} \right)\ .
\label{100}
\end{equation}
It is remarkable, that
the matrix $ \hat{\bf B}$, different from ${\bf B}$, does {\it not} depend on retarded time.
This property allows us to find the
solution of equation
(\ref{95}) in terms of elementary functions as
\begin{equation}
\left(\begin{array}{cc}
Y_2(u)  \\
Y_3(u)
\end{array} \right)
=
{\bf R}(u) \cdot
\left(\begin{array}{cc}
Y_2(0)  \\
Y_3(0)
\end{array} \right) , \quad
{\bf R}(u)
\equiv
\left(\begin{array}{cc}
\cos\Phi(u)  & - \sin\Phi(u)  \\
\sin\Phi(u)  &  \cos\Phi(u)
\end{array} \right) \ ,
\label{101}
\end{equation}
where
\begin{equation}
\Phi(u) \equiv  \Pi \int\limits_0^u \frac{du}{L^2(u)} \ \  .
\label{102}
\end{equation}
The combination
\begin{equation}
Y^2_2(u) + Y^2_3(u) = Y^2_2(0) + Y^2_3(0)
\label{103}
\end{equation}
is preserved, i.e. the dynamics of ${\bf Y}$ represents a rotation in the $x^2 0 x^3$ plane.
The functions $S_2(u)$ and $S_3(u)$ in (\ref{86})) can now be expressed in terms of the three matrices
${\bf H}(u)$, ${\bf T}(u)$ and ${\bf R}(u)$, given by the expressions (\ref{71}), (\ref{97})
and (\ref{101}), respectively:
\begin{equation}
\left(\begin{array}{cc}
S_2(u)  \\
S_3(u)
\end{array} \right)
{=}
{\bf T}(u) \cdot  {\bf R}(u) \cdot
\left(\begin{array}{cc}
S_2(0)  \\
S_3(0)
\end{array} \right)
{+} \frac{E_v}{C_v} \left[ \ {\bf H}(u) {-} {\bf T}(u) \cdot {\bf R}(u)
\right]
\left(\begin{array}{cc}
C_2  \\
C_3
\end{array} \right)\ .
\label{104}
\end{equation}
While the matrices ${\bf T}(u)$ and ${\bf R}(u)$ are constructed
out of elementary functions, the matrix ${\bf H}(u)$, according to
Eq.~(\ref{71}), consists of Hill functions, which for the special
case of subsection \ref{sandwich} reduce to Mathieu functions. The
latter, in turn, can be expressed via ``cosine elliptic'' and
``sine elliptic'' functions (see the discussion following
Eqs.~(\ref{77})). All these functions are assumed to be known
here. In the absence of the GW,
\begin{equation}
{\bf T}(u) \equiv {\bf I}, \quad
{\bf H}(u) \ = \ {\bf R}(u) \ \equiv {\bf R}_0(\tau) =
\left(\begin{array}{cc}
\cos\Omega_H\tau  & - \sin\Omega_H\tau  \\
\sin\Omega_H\tau  &  \cos\Omega_H\tau
\end{array} \right) ,
\label{105}
\end{equation}
and we recover the standard flat spacetime rotation of the spin particle,
\begin{equation}
\left(\begin{array}{cc}
p_2(\tau)  \\
p_3(\tau)
\end{array} \right)
=
{\bf R}_0(\tau) \cdot
\left(\begin{array}{cc}
C_2  \\
C_3
\end{array} \right) , \quad
\left(\begin{array}{cc}
S_2(\tau)  \\
S_3(\tau)
\end{array} \right)
=
{\bf R}_0(\tau) \cdot
\left(\begin{array}{cc}
S_2(0)  \\
S_3(0)
\end{array} \right) .
\label{106}
\end{equation}
Equation (\ref{104}) represents the general solution for the spin dynamics
in the GW field
(\ref{30}) and the magnetic field (\ref{48}).
The structure of the solution (\ref{104}) allows us to interpret the spin
dynamics as
composed of three separate contributions, characterized by the matrices
${\bf R}(u)$, ${\bf T}(u)$ and ${\bf H}(u)$.
The matrix ${\bf R}$ represents a Larmor type precession with the frequency
$\frac{\Pi}{L^2(u)}$.
As already mentioned, the matrix ${\bf T}$ describes a gravitationally
induced rotation
with phase $\psi(u)$ and frequency $\dot{\psi}(u)$, and
finally, the matrix ${\bf H}$ accounts for the coupling of the particle motion
(\ref{70}) to the spin dynamics.

\section{Isospin evolution}
\label{isospin}

In this section as well as in the subsequent one we discuss the
dynamics of non-Abelian charges under the influence of external
gravitational and Yang-Mills fields. Let us consider Eq.
(\ref{20}) for the $SU(2)$ symmetry group. Since the structure
constants for this group coincide with the Levi-Civita symbol, all
three directions in the isospin space are equivalent. With the
choice $ I^{(3)}= I_{(A)}q^{(A)} $ we obtain  the following
equations for the isospin evolution:
\begin{equation}
\frac{\mbox{d}I^{(1)}}{\mbox{d}\tau}=\Omega \cdot I^{(2)}\ , \quad
\frac{\mbox{d}I^{(2)}}{\mbox{d}\tau}=-\Omega \cdot I^{(1)}\ .
\label{107}
\end{equation}
The precession frequency $\Omega $ in Eq. (\ref{20}) is calculated on
the particle
worldline with $p^i$ from (\ref{70}) and
the potential
\begin{equation}
A_i(u) {=} \frac{1}{2} M \left[\left(x^2(u) {-} x^2(0) {-}
\frac{C_2}{\Pi C_v}
\right) \delta^3_i {-} \left(x^3(u) {-} x^3(0) {-}
\frac{C_3 }{\Pi C_v} \right)
\delta^2_i \right]\ ,
\label{108}
\end{equation}
corresponding to the constant solution $F_{23} = M $.
Differentiating the expression (\ref{108}), we recover the field strength
(\ref{49}).
The frequency $\Omega(u)$ in (\ref{20})  is given by
\begin{equation}
\Omega(u) = \frac{g}{m} (A_2 p^2 + A_3 p^3)\ .
\label{109}
\end{equation}
Here, the arbitrary constant was chosen such that $\Omega(u=0) = 0$.
In order to find the terms $x^2(u) {-} x^2(0)$ and $x^3(u) {-} x^3(0)$ which are needed
in (\ref{108}), we have to integrate the second
equation in (\ref{35}). The formal solution is
\begin{equation}
x^{\alpha}(u) {-} x^{\alpha}(0) = \frac{1}{C_v} \int_0^u d\xi g^{\alpha
\beta}(\xi) p_{\beta}(\xi) \ ,
\label{110}
\end{equation}
which provides us with
\begin{equation}
\Omega(u)= \frac{gM}{2mC_v}
\left\{
\int_{0}^{u}\mbox{d}\xi \left[p^2(\xi) p^{3}(u) - p^3(\xi) p^2(u) \right] -
\frac{1}{\Pi} \left[ C_2 p^3(u) - C_3 p^2(u) \right] \right\}\ .
\label{111}
\end{equation}
Again we assume here $p^2(u)$ and $p^3(u)$ to be known, i.e., the
particle dynamics is considered to be solved [cf, Eqs.~(\ref{70})
and (\ref{71})].  The solution of the system (\ref{107}) then
becomes
\begin{equation}
I^{(1)} = I \cos\Psi(u)\ , \quad I^{(2)}= - I \sin\Psi(u)\ , \quad
\Psi(u) = \Psi(0) + \frac{m}{C_{v}} \int_{0}^{u} \Omega(u) \mbox{d}u\ .
\label{112}
\end{equation}
The function $\Psi(u)$ plays the role of the (generally $u$-dependent)
phase of the isospin precession in the external Yang-Mills field \cite{BaSu}.
The set of equations (\ref{112}) with (\ref{111}) provides a complete
description
for the isospin dynamics under the influence of the GW (\ref{30}) and the
Yang-Mills field (\ref{49}).

\section{Color charge evolution}
\label{color}

The SU(3) case may be studied along similar lines although it is technically
more
extended since more degrees of freedom are involved.
As a result, we shall find a richer dynamical structure than in the SU(2)
case.
With the ansatz (\ref{18}) and the expressions (\ref{15}), Eqs. (\ref{24})
for the color
charge dynamics become
\begin{eqnarray}
\frac{\mbox{d}Q^{(1)}}{\mbox{d}\tau}&{=}&{-}\Omega
\left[(q^{(2)}Q^{(3)}
{-}q^{(3)}Q^{(2)}){+}\frac{1}{2}(q^{(4)}Q^{(7)}{-}q^{(7)}Q^{(4)}){-}
\frac{1}{2}(q^
{(5)}Q^{(6)}{-}q^{(6)}Q^{(5)})\right]\ ,\nonumber\\
\frac{\mbox{d}Q^{(2)}}{\mbox{d} \tau}&{=}&{-}\Omega
\left[(q^{(3)}Q^{(1)}{-}q^{(1)}Q^{(3)}){+}\frac{1}{2}(q^{(4)}Q^{(6)}
{-}q^{(6)}Q^{(4)}){+}\frac{1}{2}(q^{(5)}Q^{(7)}{-}q^{(7)}Q^{(5)})\right]\
,\nonumber\\
 \frac{\mbox{d}Q^{(3)}}{\mbox{d}
\tau}&{=}&{-}\Omega\left[(q^{(1)}Q^{(2)}{-}q^{(2)}
Q^{(1)}){+}\frac{1}{2}(q^{(4)}Q^{(5)}{-}q^{(5)}Q^{(4)}){-}\frac{1}{2}
(q^{(6)}Q^{(7)}{-}q^{(7)}Q^{(6)})\right]\ ,\nonumber\\
\frac{\mbox{d}Q^{(4)}}{\mbox{d}\tau}&{=}&{-}\frac{\Omega}{2}
\left[(q^{(7)}Q^{(1)}
{-}q^{(1)}Q^{(7)}){+}(q^{(6)}Q^{(2)}{-}q^{(2)}Q^{(6)})
{+}(q^{(5)}Q^{(3)}{-}q^{(3)}Q^{(5)})\right]\nonumber\\
&&-\frac{\sqrt{3}\Omega}{2}[q^{(5)}Q^{(8)}-q^{(8)}Q^{(5)}] \ ,\nonumber\\
\frac{\mbox{d}Q^{(5)}}{\mbox{d}\tau}&{=}&{-}\frac{\Omega}{2}
\left[(q^{(1)}Q^{(6)}{-}q^{(6)}
 Q^{(1)}){+}(q^{(7)}Q^{(2)}{-}q^{(2)}Q^{(7)})
{+}(q^{(3)}Q^{(4)}{-}q^{(4)}Q^{(3)})\right]\nonumber\\
&&{-}\frac{\sqrt{3}\Omega}{2}[q^{(8)}Q^{(4)}{-}q^{(4)}Q^{(8)}]\ ,\nonumber\\
\frac{\mbox{d}Q^{(6)}}{\mbox{d}\tau}&{=}&{-}\frac{\Omega}{2}
\left[(q^{(5)}Q^{(1)}{-}
q^{(1)}Q^{(5)}){+}(q^{(2)}Q^{(4)}{-}q^{(4)}Q^{(2)}){+}(q^
{(3)}Q^{(7)}{-}q^{(7)}Q^{(3)})\right]\nonumber\\
&&{-}\frac{\sqrt{3}\Omega}{2}[q^{(7)}Q^{(8)}{-}q^{(8)}Q^{(7)}]\ ,\nonumber\\
\frac{\mbox{d}Q^{(7)}}{\mbox{d}\tau}&{=}&{-}\frac{\Omega}{2}
\left[(q^{(1)}Q^{(4)}{-}q^{(4)}Q^{(1)}){+}(q^{(2)}Q^{(5)}{-}q^{(5)}Q^{(2)})
{+}(q^{(6)}Q^{(3)}{-}q^{(3)}Q^{(6)})\right]\nonumber\\
&&{-}\frac{\sqrt{3}\Omega}{2}[q^{(8)}Q^{(6)}{-}q^{(6)}Q^{(8)}]\ ,\nonumber\\
\frac{\mbox{d}Q^{(8)}}{\mbox{d}\tau}&{=}&{-}\frac{\sqrt{3}\Omega}{2}
\left[(q^{(4)}
Q^{(5)}{-}q^{(5)}Q^{(4)}){+}(q^{(6)}Q^{(7)}{-}q^{(7)}Q^{(6)})\right]\ .
\label{113}
\end{eqnarray}
The space of color charges may be split into three different subspaces,
which correspond to the structures of the $SU(2)$, $SU(2) \times
U(1)$ and $U(1)$ subgroups  of the total group $SU(3)$ (see, e.g.
\cite{ElDa}). In the following subsections we consider the vector $q^{(A)}$ to lie in the first, second, and third subspaces, respectively.

\subsection{First special case}

Let the vector $q^{(A)}$ have only the three non-zero components  $q^{(1)}$,
$q^{(2)}$ and $q^{(3)}$.
It is then evident that
\begin{equation}
(Q^{(1)})^2  + (Q^{(2)})^2 + (Q^{(3)})^2 = {\rm const}\ ,
\label{114}
\end{equation}
\begin{equation}
(Q^{(4)})^2 + (Q^{(5)})^2 + (Q^{(7)})^2 + (Q^{(7)})^2 = {\rm const}\ ,
\label{115}
\end{equation}
and
\begin{equation}
Q^{(8)} = {\rm const}\ .
\label{116}
\end{equation}
For the color charges $Q^{(1)}$, $Q^{(2)}$, and $Q^{(3)}$,
which correspond to a  $SU(2)$ subgroup  of the total $SU(3)$ group,
the combination (\ref{114}) is preserved.
A similar relation holds for the set
$Q^{(4)}$, $Q^{(5)}$, $Q^{(6)}$, and $Q^{(7)}$, while $Q^{(8)}$ is
separately conserved.
Relations (\ref{114})-(\ref{116})  are also obtained for the case that the only
non-vanishing components are $q^{(4)}$, $q^{(5)}$, $q^{(6)}$, and $q^{(7)}$,
as well as for
the choice
$q ^{\left(1 \right)}=q ^{\left(2 \right)}=.....=q
^{\left(7 \right)}=0$ and $q ^{\left(8 \right)}\neq 0$.
Let us now further specify to the case $q^{(A)}=\delta^{(A)}_{(1)}$.
Then the system (\ref{113})  takes the form
$$
\frac{\mbox{d}Q^{(1)}}{\mbox{d} \tau}=0\ , \quad
\frac{\mbox{d}Q^{(8)}}{\mbox{d} \tau}=0\ , \quad
$$
\begin{eqnarray}
\frac{\mbox{d}Q^{(2)}}{\mbox{d} \tau}=\Omega Q^{(3)}\ ,&&\quad
\frac{\mbox{d}Q^{(3)}}{\mbox{d} \tau}=-\Omega Q^{(2)}\ ,\nonumber\\
\frac{\mbox{d}Q^{(4)}}{\mbox{d} \tau}=\frac{1}{2}\Omega Q^{(7)}\ ,&&\quad
\frac{\mbox{d}Q^{(7)}}{\mbox{d} \tau}=-\frac{1}{2}\Omega Q^{(4)}\ ,\nonumber\\
\frac{\mbox{d}Q^{(5)}}{\mbox{d} \tau}=-\frac{1}{2}\Omega Q^{(6)}\ ,&&\quad
\frac{\mbox{d}Q^{(6)}}{\mbox{d} \tau}=\frac{1}{2}\Omega Q^{(5)}\ .
\label{117}
\end{eqnarray}
The color charge $Q^{(1)}$ remains constant because it is the projection
$Q^{(A)} q_{(A)}$ of the vector $Q^{(A)}$ on the given preferred direction
$q_{(A)}$.
The charge $Q^{(8)}$ does not evolve, because for such a
$q_{(A)}$ the antisymmetric tensor $H_{(A)(B)}$ in (\ref{24}) does not contain
a non-vanishing component with $(A)=(8)$.
The equations for $Q^{(2)}$, ...
$Q^{(7)}$ split into {\it three
two-dimensional} subsystems with the
pairs $Q^{(2)}$ and $Q^{(3)}$,
$Q^{(4)}$ and $Q^{(7)}$, $Q^{(5)}$ and
$Q^{(6)}$. The evolution of the first
pair ($Q^{(2)}$ and $Q^{(3)}$)
corresponds to a
precession in the group space with
the
frequency $\Omega$.
It has a solution of the type (\ref{112}).
The dynamics of the pairs $Q^{(4)}$, $Q^{(7)}$ and
$Q^{(5)}$, $Q^{(6)}$ is a precession with
the frequency $\Omega/2$.

\subsection {Second special case: $q^{(A)}=\delta^{(A)}_{(4)}$}

Now we assume $q^{(A)}$ to lie in the second subspace. As an example we consider the case $q^{(A)}=\delta^{(A)}_{(4)}$.
For this choice the set of equations (\ref{113})  can be transformed into
$$
\frac{\mbox{d}Q^{(4)}}{\mbox{d} \tau}=0, \quad
\frac{\mbox{d}}{\mbox{d} \tau}\left(- \frac{\sqrt3}{2}Q^3 + \frac{1}{2}Q^8 \right) = 0\ ,
$$
$$
\frac{\mbox{d}Q^{(5)}}{\mbox{d} \tau}=\Omega Q^*\ , \quad
\frac{\mbox{d} Q^*}{\mbox{d}\tau}=-\Omega Q^{(5)}\ , \quad
Q^*  \equiv \frac{1}{2}(Q^{(3)}+\sqrt{3}Q^{(8)})\ ,
$$
\begin{eqnarray}
\frac{\mbox{d}Q^{(1)}}{\mbox{d} \tau}&=&-\frac{1}{2}\Omega Q^{(7)}\ , \quad
\frac{\mbox{d}Q^{(7)}}{\mbox{d} \tau}=\frac{1}{2}\Omega Q^{(1)}\ ,\nonumber\\
\frac{\mbox{d}Q^{(2)}}{\mbox{d}\tau}&=&-\frac{1}{2}\Omega Q^{(6)}, \quad
\frac{\mbox{d}Q^{(6)}}{\mbox{d} \tau}=\frac{1}{2}\Omega Q^{(2)}\ .
\label{118}
\end{eqnarray}
The quantities $Q^{(4)}$ and
$- \frac{\sqrt3}{2}Q^3 + \frac{1}{2}Q^8$ remain constant, the projections
$Q^{(5)}$ and $Q^*  \equiv \frac{1}{2}(Q^{(3)}+\sqrt{3}Q^{(8)})$ precess with
the frequency $\Omega$, the pairs $Q^{(1)}$, $Q^{(7)}$ and  $Q^{(2)}$,
$Q^{(6)}$ precess with $\Omega/2$.
Furthermore, one has
$$
(Q^{(1)})^2 + (Q^{(7)})^2 = {\rm const}\ , \quad
(Q^{(2)})^2 + (Q^{(6)})^2 = {\rm const}\ ,
$$
\begin{equation}
(Q^{(3)})^2 + (Q^{(5)})^2 + (Q^{(8)})^2 = {\rm const}\ .
\label{119}
\end{equation}

\subsection{Third special case: $q^{(A)}=\delta^{(A)}_{(8)}$}

Finally, let $q^{(A)}$ be parallel to the basis vector in the $U(1)$
subspace, i.e.,
$q^{(A)}=\delta^{(A)}_{(8)}$.
Here we obtain
$$
\frac{\mbox{d}Q^{(1)}}{\mbox{d} \tau}=0\ , \quad
\frac{\mbox{d}Q^{(2)}}{\mbox{d}\tau}=0\ , \quad
\frac{\mbox{d}Q^{(3)}}{\mbox{d}\tau}=0\ ,
\quad
\frac{\mbox{d}Q^{(8)}}{\mbox{d} \tau}=0\ , \quad
$$
as well as
\begin{eqnarray}
\frac{\mbox{d}Q^{(4)}}{\mbox{d} \tau}&=&\frac{\sqrt{3}}{2}\Omega Q^{(5)}\ ,
\quad
\frac{\mbox{d}Q^{(5)}}{\mbox{d} \tau}=-\frac{\sqrt{3}}{2}\Omega Q^{(4)}\ ,
\nonumber\\
\frac{\mbox{d}Q^{(6)}}{\mbox{d} \tau}&=&\frac{\sqrt{3}}{2}\Omega Q^{(7)}\ ,
\quad
\frac{\mbox{d}Q^{(7)}}{\mbox{d}\tau}=-\frac{\sqrt{3}}{2}\Omega Q^{(6)}\ .
\label{120}
\end{eqnarray}
The pairs
$Q^{(4)}$, $Q^{(5)}$ and $Q^{(6)}$, $Q^{(7)}$ precess with the frequency
$\frac{\sqrt{3}}{2}\Omega$, while
$Q^{(1)}$, $Q^{(2)}$, $Q^{(3)}$, and $Q^{(8)}$ are constant.

\subsection{Remarks on the general case}

In the general case one expects the color vector $Q^{(A)}$ to rotate
in the hypersurface
orthogonal to $q^{(A)}$ in the group space. This is illustrated by the
following analogy.
Let us consider the standard
decomposition of the Maxwell tensor with respect to a four-velocity
vector $V^i$
of an arbitrary observer,
\begin{equation}
F_{ik} = E_i V_k - E_k V_i - \epsilon_{ikjl} H^j V^l \ ,
\label{121}
\end{equation}
where $E^i$ and $H^k$ are the corresponding
four-vectors for the electric and magnetic fields, respectively.
For $F_{ik}V^k = 0$ the comoving observer experiences a magnetic field
only which, via the Lorentz force
generates a spatial
particle rotation, i.e. a rotation in the hypersurface
orthogonal to $V^i$.

In the present case the vector $q^{(A)}$ plays the role of $V^i$.
Instead of Maxwell's
tensor we have to consider [cf. Eq.~(\ref{24})] the antisymmetric tensor
$H_{(A)(C)}= f_{(A)(B)(C)} q^{(B)}$.
Since $H_{(A)(C)} q^{(C)} = 0$, this represents a group space analogue to
the previous case of
a pure magnetic field in spacetime. Consequently, the color vector rotates
in the group space
analogously to the momentum vector of a charged particle in a pure magnetic
field.

\section{Conclusions}
\label{conclusions}

Charged particles in electromagnetic fields are known to be parametrically
influenced by gravitational waves.
Typical phenomena are parametric resonances and parametric oscillations.
In the present paper we have generalized and extended work in this field to
include
classical spin particles and particles with non-Abelian charges in specific
Yang-Mills fields.
Moreover, no weak field approximation was used for the GW.
The electrically charged spin particle was described by the
Bargmann-Michel-Telegdi equations.
For the dynamics of the non-Abelian charges we used Wong's equations for the
isospin (SU(2)-symmetry) and for the color charges (SU(3)-symmetry).
We derived exact general solutions for the parametric influence of the
GW on the particle motion in each of
the mentioned cases, including the dynamics of spin, isospin, and color
charge.
For the case of a special sandwich GW the particle dynamics was
reduced to a set of Mathieu equations.
Using well-known stability properties
of the latter we found that parametric instabilities are a generic phenomenon
for such kind of particle motion.
Since spin, isospin and color charge are
coupled to this motion, their dynamics is affected correspondingly.
The spin
dynamics was shown to be composed of three elements, namely a gravitationally
modified Larmor precession, a part due to the coupling to the particle motion,
and a pure gravitational part.
The vectors of isospin and color charge carry
out gravitationally influenced precession motions in their group spaces which
we have classified for several cases.\\  \ \\

{\bf Acknowledgments}\\
 \ \\
 This paper was supported by the Deutsche
Forschungsgemeinschaft and NATO grant PST. CLG.977973.
The authors thank F.G. Suslikov for useful discussions about the SU(3)
symmetry.

\end{document}